\documentclass[aps,showpacs]{revtex4}
\usepackage{amsbsy}
\usepackage{graphicx,color}
\usepackage{amsfonts}
\usepackage{amsmath}
\usepackage{amssymb}

\newcommand{\ds}{ _{\downarrow}}
\newcommand{\us}{ _{\uparrow}}
\newcommand{\up}{\uparrow}
\newcommand{\down}{\downarrow}

\begin{document}
\draft
\title{Atom-dimer scattering amplitude for fermionic mixtures with different masses: s-wave and p-wave contributions}
\author{F. Alzetto$^{(a)}$, R. Combescot$^{(a),(b)}$ and  X. Leyronas$^{(a)}$}
\address{(a) Laboratoire de Physique Statistique, Ecole Normale Sup\'erieure, UPMC  
Paris 06, Universit\'e Paris Diderot, CNRS, 24 rue Lhomond, 75005 Paris,  
France.}
\address{(b) Institut Universitaire de France,
103 boulevard Saint-Michel, 75005 Paris, France.}
\date{Received \today}
\pacs{03.75.Ss, 03.65.Nk, 34.50.Cx, 67.85.Lm}

\begin{abstract}
We study near a Feshbach resonance, as a function of the mass ratio, the fermion-dimer scattering amplitude 
in fermionic mixtures of two fermion species. When masses are equal the physical situation is known to be quite simple. 
We show that, when the mass ratio is increased, the situation becomes much more complex.
For the s-wave contribution we obtain an analytical solution in the asymptotic limit of very large mass ratio.
In this regime the s-wave scattering amplitude displays a large number of zeros, essentially linked to
the known large value of the fermion-dimer scattering length in this regime. We find by an exact numerical calculation
that a zero is still present for a mass ratio of 15. For the p-wave contribution we make our study below the mass ratio
of 8.17, where a fermion-dimer bound state appears. We find that a strong p-wave resonance is present at low energy,
due to a virtual bound state, in the fermion-dimer system, which is a forerunner of the real bound state. This resonance
becomes prominent in the mass ratio range around the one corresponding to the $^{40}$K - $^6$Li mixtures, much 
studied experimentally. This resonance should affect a number of physical properties. 
These include the equation of state of unbalanced mixtures at very low temperature but also the
equation of state of balanced mixtures at moderate or high temperature. The frequency and the damping of collective modes
should also provide a convenient way to evidence this resonance. Finally it should be possible to modify the effective mass
of one of the fermionic species by making use of an optical lattice. This would allow to study the strong dependence of the
resonance as a function of the mass ratio of the two fermionic elements. In particular one could check if the virtual bound state
is relevant for the instabilities of these mixtures.
\end{abstract}
\maketitle

\section{INTRODUCTION}

One of the most fascinating aspects of the physics of ultracold gases \cite{gps} is the ability, 
in appropriate situations, to describe the interaction between non identical fermionic atoms
in terms of a single parameter, namely the scattering length $a$ for two different atoms.
For example in fermionic gases involving only atoms in the two lowest hyperfine states of a single element,
such as $^6$Li or $^{40}$K, the interaction is fully described by the scattering length between atoms belonging to these two
different hyperfine states \cite{gps}. Moreover in the presence of a Feshbach resonance this
scattering length can be experimentally modified almost at will by merely changing the applied magnetic field. The case of wide Feshbach
resonances is particularly convenient since all the physics related to the origin of the resonance, namely the existence of
a closed channel, is irrelevant for most purposes. This has allowed, in the case of two fermionic atomic species, the experimental 
realization of the BEC-BCS crossover where the scattering length goes from small negative values to small positive values through 
the resonance, where it diverges. For small negative values of the scattering length two different atoms have a weak effective attraction, 
which for high enough density and small enough temperature, leads to the formation of Cooper pairs analogous to the ones appearing
in superconductors and corresponding superfluid properties. On the other hand for small positive values of the scattering length $a>0$,
two fermionic atoms of different species form a bound state resulting in the appearance of molecules, or dimers, 
with binding energy $E_b$.
These dimers behave statistically as bosons, and again for high enough density and small enough temperature a Bose-Einstein
condensate forms with superfluid properties. The BEC-BCS crossover occurs when one goes continuously from the Bose
condensate to the BCS superfluid by merely varying the scattering length through the change of magnetic field.

The simplicity of the description of the interaction extends to the BEC limit, when one considers dimer-dimer interaction, while this is a priori a more complex situation since dimers have an internal structure.
Nevertheless the BEC limit is quite simple since the composite bosons one deals with are dilute. 
In this case again the interaction between these bosons is fully characterized by their scattering length $a_4$. 
This is because in the low temperature regime corresponding to the existence of the Bose-Einstein condensate,
the kinetic energies of the dimers will be quite small compared to their binding energy, and the structure of these composite bosons will be irrelevant. This scattering length $a_4$ has been obtained \cite{pss,bkkcl}, in terms of the scattering length $a$ for fermions, as
$a_4 \simeq 0.6 \, a$ when the different fermions have equal masses. 

On the other hand when one goes away from the BEC limit toward
the unitarity limit, it is no longer true that the composite nature of the bosons is irrelevant.
One might accordingly expect a more complex physics for this molecular gas. 
Remarkably the $T=0$ equation of state on this BEC
side is nevertheless very well described \cite{gps} in a wide domain of density by the Lee-Huang-Yang \cite{lhy} equation of state
for these composite bosons \cite{lc}, making again merely use of the dimer-dimer scattering length $a_4$
between these bosons. This is as if these bosons were elementary. This is somewhat surprising.
 
More precisely, assuming for example two fermions species with equal densities $n=k_F^3/6\pi ^2$, a typical
fermion wavevector is of order $k_F$, and for $k_Fa \sim 1$ the typical kinetic energy of a fermion is now of the same order as the
dimer binding energy which is $E_b=1/(2\mu a^2)$
with $\mu$ being the reduced mass.
In this case one has to consider the whole
energy dependent scattering amplitude and not only its zero energy limit given by the scattering length. This is valid for dimer-dimer
scattering as well as for single fermion-dimer scattering. In this last case these single fermions may arise because of thermal breaking
of dimers, or because we consider unbalanced fermionic mixtures. Actually for $k_Fa \sim 1$ these single fermions arise also because we
may no longer consider isolated dimers since they are too close, and we have actually to deal with a complicated many-body problem.
However our point is that even the simplest minded description has to consider the energy dependence of scattering processes.

Actually it is possible to understand qualitatively the wide range of validity of the Lee-Huang-Yang equation of state 
in the following way. One assumes first that only the s-wave scattering properties are relevant, since higher partial wave
give a quite small contribution at low energy. Then one supposes that the s-wave phase shift $\delta_0(k)$ stays small enough so that 
its low energy expansion $\delta_0(k) \simeq -k{\tilde a}$ remains valid, where ${\tilde a}$ is the scattering length relevant for the considered process, dimer-dimer or fermion-dimer scattering.
In this case the standard low energy expression for the scattering amplitude $f(k)=-1/({\tilde a}^{-1}+ik)$ remains valid.
This is what happens when the different fermions have equal masses. In this case the
scattering amplitude has a fairly weak variation when $k$ is going from zero to the maximum value corresponding to dimer breaking, so the situation is not much different from evaluating the scattering amplitude at $k=0$.

The purpose of the present paper is to show
that this simplicity is  linked to the fact that the two atomic species have equal masses $m\us=m\ds$
(we follow the standard use of referring to the two different atomic species as $\up$ and $\down$ atoms, even if there is
no relation between this notation and the physical spin of these atoms).
It disappears when the masses are quite different. In this case, even though the elementary scattering process between different
fermions is very simple to describe, nevertheless the mere complexity due to the existence of a dimer leads to complexity
in the involved scattering processes. In the present paper we will consider only the simplest situation which displays such a feature.
We will deal with the fermion-dimer scattering, not the dimer-dimer scattering. The lonely fermion has an $\up$ spin
and scatters with a $\up$-$\down$ dimer.  We will furthermore assume that the available kinetic energy
is small enough so the dimer is not broken by the collision. In other words we will deal only with elastic scattering.

In a first part of the paper
we will consider s-wave scattering, since it is the dominant one at very low energy and is anyway expected to
be an essential component of the scattering process. This is the natural extension of the case where the
scattering is fully described by the scattering length.
Actually, although we have performed numerical calculations for any value of the mass ratio $m\us/m\ds$, we deal
extensively with the limit where this mass 
ratio is large. Indeed in this case we will be able to obtain an analytical solution of the problem, which has the advantage of displaying quite explicitly
the behaviour resulting from the mass difference between the two fermion species. Naturally since we have already
explored in a preceding work \cite{acl} the behaviour of the scattering length in this limit, this paper can also be seen as a natural
extension of our earlier work. But it has the major interest of displaying a strong structure in the s-wave scattering
amplitude, which is actually essentially linked to the increase of the scattering length with the mass ratio \cite{ism,isk,acl}.

Then in the second part of our paper we will turn to p-wave scattering. 
Although it does not contribute in the zero energy limit,
this partial wave turns out to be quite important. This is physically linked to the appearance of new bound states which 
appear at the level of the three-body problem when the mass ratio increases.
These bound states have been
studied in great details by Kartavtsev and Malykh \cite{karma}, for the present situation of 
two identical fermions with mass $m\us$ and a different fermion
with mass $m\ds$. They are not present in the s-wave channel and appear only
for angular momenta $\ell \ge 1$. Among them are the well-known Efimov \cite{efimov} states, which are quite remarkable for
their spectrum structure with binding energy going up to infinity when the range of the interaction goes to zero.
For $\ell =1$ the Efimov states appear only \cite{efi1,karma} for a mass ratio $m\up / m\down = 13.6$. However there are already bound states
for smaller mass ratios, which are not Efimov states. A first bound state \cite{karma} appears for mass ratio $m\up / m\down = 8.17$.

Actually very important contributions in the fermion-dimer scattering properties arise even for quite lower mass ratios.
The basic reason is that, even before the mass ratio threshold for bound states is reached, virtual bound states are already present.
For nonzero angular momenta, in particular $\ell=1$, the centrifugal barrier acts to inhibit their decay and provide them with
fairly long lifetime. Hence they are not much different from real bound states. Moreover, since their energy is positive, they
give rise to resonances, which are strong when they are located at low energy. 
In particular the p-wave contribution becomes very strong, due to quasi-resonance, when the mass ratio approaches $8.17$.
However we find that these effects are already important
for the mass ratio $m\up / m\down = 6.64$ corresponding to the mixtures of $^6$Li and $^{40}$K, which is probably 
presently the most
investigated experimentally\cite{munich,innsb,ens} and which is of very high current interest
\cite{pao,parish,blume,baranov,ops,brs, diener,ggsc,bdgc}. 
Actually this point has already been noted, for the specific case of the $^6$Li-$^{40}$K mixtures, 
by Levinsen \emph{et al} \cite{lp}. This was mostly done for the case of narrow Feshbach resonances, which are by far
the most frequently found in these mixtures. They were also mostly interested in lower dimension physics.

Here we will consider only the case of wide Feshbach resonance in 3D, because of its theoretical interest due to its
simplicity.
In the following we will assume that we are near such a Feshbach resonance and ignore any stability problem.
Fortunately such a Feshbach resonance, which is fairly broad and reasonably stable, has been identified 
experimentally near $155$G \cite{inns} in $^6$Li-$^{40}$K mixtures.
We will consider in detail the mass ratio dependence of the p-wave
contribution. This is not only of theoretical interest since we will show that, by making use of optical lattices,
it is possible to vary experimentally the effective mass ratio of $^6$Li-$^{40}$K mixtures. This would allow to
study experimentally the effect of the p-wave contribution when it is quite strong, or conversely to eliminate it
to a large extent which could be in particular be quite useful if this contribution turned out to be experimentally
detrimental. We note indeed that these virtual bound states could play a very important role in the three-body decay
and in the stability of these mixtures, since they correspond to physical situations where three atoms stay close together
for a fairly long time, increasing in this way the probability of three-body decay processes. In this way optical lattices 
could also be used to study and possibly remove instability problems.

In this  p-wave study we will consider mass ratios below the value $8.17$, corresponding to the appearance of the first p-wave bound state, since clearly beyond this threshold the physical situation will get more complicated.
This justifies also that we consider only the s-wave and p-wave components of the full scattering amplitude,
since the higher angular momentum components are expected to give very small contributions
to the total scattering in the regime we investigate.

In practice the next section \ref{swave} is devoted to the study of the s-wave contribution to the scattering amplitude. But in the first subsection
we will present the basic general equations which will be used also for the p-wave contribution. Then in the rest of this next section \ref{swave} we will
study analytically in details the case where the mass ratio $m\up / m\down$ is very large. We will conclude this section by considering
numerically cases where this mass ratio corresponds to situations which can be reached experimentally. Then in the last section \ref{pwav}
before our conclusion we study numerically the p-wave contribution.

\section{s - wave scattering}\label{swave}

\subsection{Basic equation}\label{base}

In the same way as in our earlier work \cite{acl} we use the integral equation formulation of this dimer-fermion scattering problem first worked out by 
Skorniakov and Ter-Martirosian\cite{stm}. It is convenient to derive it rapidly in the case where the fermion masses are different \cite{sademelo}
by writing directly \cite{bkkcl} the integral equation satisfied by the full dimer-fermion scattering vertex $T_3(p_1,p_2;P)$.
Here $P$ is the (conserved) momentum-energy of the particles $P \equiv \{\textbf{P},E\}$, with
$\textbf{P}$ the total momentum and $E$ the total energy of incoming particles. $p_1$ is the momentum-energy of the incoming
$\up$ fermion and $p_2$ its outgoing value. Since we want the dimer-fermion scattering amplitude, we can take the center of
mass referential for which $\textbf{P}={\bf 0}$. If we call ${\bf k}_0$ the outgoing momentum of the $\up$ fermion, i.e. ${\bf p}_2
={\bf k}_0$, the total energy $E$ is the sum of the binding energy of the dimer and of the kinetic energies of the outgoing
particles:
\begin{eqnarray}\label{totE}
E=-\frac{1}{2\mu a^2}+\frac{k_0^2}{2\mu _T}
\end{eqnarray}
where $\mu =m\us m\ds /(m\us + m\ds)$ is the dimer reduced mass, $\mu _T$ is the atom-dimer reduced mass 
$\mu _T = m\us M /(m\us + M)=m\us (m\us + m\ds)/(2 m\us +m\ds)$.
Here $M=m\us + m\ds$ is the mass of the dimer.
The integral equation satisfied by $T_3(p_1,p_2;P)$ is \cite{bkkcl}:
\begin{equation}\label{3Vertex}
    T_3(p_1,p_2;P)=-G\ds(P-p_1-p_2)- \sum_{q}G\ds(P-p_1-q)
    G\us(q)\, T_2(P -q)\; T_3(q, p_2; P),
\end{equation}
where $\sum\limits_q\equiv i \int d{\bf q}\,d\Omega/(2\pi)^4$, with $q \equiv \{\textbf{q},\Omega \}$. Here $T_2(P)$ is the dimer propagator:
\begin{eqnarray}\label{}
    T_2(P) = \frac{2\pi}{\mu}\,
   \frac{1}{a^{-1}
    -\sqrt{2\mu (\mathbf{P}^2/2M-E})} 
\end{eqnarray}
while $G_{\up , \down}(p)$ are the single fermion $\up$ or $\down$ propagators. Just as in the equal mass case \cite{bkkcl} one can show
that, in the right-hand side of Eq.(\ref{3Vertex}), only the on-the-shell value of $T_3(q, p_2; P)$ with respect to the variable $q$
is needed. Hence we may restrict ourselves to consider only this on-the-shell value everywhere in Eq.(\ref{3Vertex}). Finally we have also to take
the on-the-shell value for $p_2 \equiv \{\textbf{p}_2,\omega_2\}$ to obtain the scattering amplitude, which means $\omega_2=k_0^2/2m\us$.
We are thus led to consider more specifically $T_3(\{\textbf{k},k^2/2m\us\},\{\textbf{k}_0,k_0^2/2m\us\};\{{\bf 0},E\})$ 
and the scattering amplitude will be obtained by taking $|{\bf k}|=k_0$. Hence we introduce, instead of $T_3(p_1,p_2;P)$, 
a function $a_3({\bf k},{\bf k}_0)$ which is directly related to the scattering amplitude, as we will see below \cite{stm}:
\begin{equation}\label{defa3}
  a_3({\bf k},{\bf k}_0) \equiv  \frac{\mu _T}{2\mu ^2a}\;\left[1+\sqrt{1+\frac{\mu a^2 }{\mu _T}(k^2-k_0^2)}\,\right]
  T_3 (\{\textbf{k},k^2/2m\us\},\{\textbf{k}_0,k_0^2/2m\us\};\{{\bf 0},E\}) 
\end{equation}

In this part we are only interested in the s-wave component of the scattering amplitude. 
This component is obtained from the above $T_3$ by averaging it over the angle
between the incoming momentum ${\bf k}$ and the outgoing one ${\bf k}_0$, and we note ${\bar T}_3$ the resulting
angular average. Noting merely $a_3(k)$ the corresponding average on $a_3({\bf k},{\bf k}_0)$ (naturally $a_3(k)$ still
depends on $k_0=|{\bf k}_0|$), we see that it is simply
given by Eq.(\ref{defa3}) where $T_3$ has been replaced by ${\bar T}_3$. Obviously when we take the angular
average of integral equation Eq.(\ref{3Vertex}), we are left with an integral equation involving only ${\bar T}_3$,
or equivalently $a_3$.

It is more convenient to write the resulting integral equation by making use of 
reduced units, obtained by setting $k={\bar k}/a$, $k_0={\bar k}_0/a$, $q={\bar q}/a$ and $a_3(k)/a={\bar a}_3({\bar k})$. We obtain:
\begin{eqnarray}\label{eqa3bar}
   R\;\frac{\bar{a}_3({\bar k})}{1+\sqrt{1+ R\,({\bar k}^2-{\bar k}_0^2)}}
   &=&\frac{1}{2R'{\bar k}{\bar k}_0}\;\ln \frac{1-R{\bar k}_0^2+{\bar k}^2+{\bar k}_0^2+R'{\bar k}{\bar k}_0}
    {1-R{\bar k}_0^2+{\bar k}^2+{\bar k}_0^2-R'{\bar k}{\bar k}_0} \\ \nonumber
   &-& \frac{1}{\pi R'} \int_{0}^{\infty} d{\bar q}\;\frac{{\bar q}^2\bar{a}_3({\bar q})}
    {\bar{q}^2-\bar{k}_0^2-i\delta}\,\frac{1}{\bar{k}\bar{q}}\;\ln \frac{1-R{\bar k}_0^2+{\bar k}^2+{\bar q}^2+R'{\bar k}{\bar q}}
    {1-R{\bar k}_0^2+{\bar k}^2+{\bar q}^2-R'{\bar k}{\bar q}}
\end{eqnarray}
where we have used for the involved mass ratios the simpler notations $R=\mu /\mu _T$, $R'=2 \mu /m\ds$.
In writing this equation we have used the fact that we consider only the case where the incoming dimer and fermion
have kinetic energies low enough so that the dimer can not be broken in the scattering process. This implies that $E<0$,
or $R{\bar k}_0^2 < 1$ in our reduced units. This makes all the arguments in the logarithms positive. Otherwise we should have
included $-i\delta$ contributions in these arguments. In the case where ${\bar k}_0 \rightarrow 0$ one checks easily
that this equation reduces to the one studied in \cite{acl}, which gives the scattering length for the same process.
It is clear that $\bar{a}_3({\bar k})$ is a complex quantity. From Eq.(\ref{eqa3bar}) one can write the coupled integral
equations for its real and imaginary parts, which are convenient for the numerical solution of Eq.(\ref{eqa3bar}).
In the case where the masses are equal $m\us = m\ds$ one can check that the resulting equations are
identical to the ones obtained by Skorniakov and Ter-Martirosian\cite{stm} for this specific case.

Finally it is convenient to consider the relation between $a_3({\bf k},{\bf k}_0)$ and the wave function 
$\psi_{\bf k}({\bf r})$ for the dimer-fermion scattering problem, as it is found in \cite{stm} and \cite{LLMQ}. 
For large dimer-fermion distance ${\bf r}$ it behaves as:
\begin{eqnarray}\label{wavef}
\psi_{\bf k}({\bf r}) \simeq e^{i{\bf k}\cdot {\bf r}}+f_k(\theta)\frac{e^{i{kr}}}{r}
\end{eqnarray}
where $\theta$ is the angle between the incoming fermion momentum ${\bf k}$ and its outgoing value ${\bf k}_0$
(with $|{\bf k}|=|{\bf k}_0|$), and $f_k(\theta)$ is the scattering amplitude.
Its Fourier transform reads:
\begin{eqnarray}\label{TF}
{\tilde \psi}_{\bf k}({\bf q})= \int d{\bf r}\,e^{-i{\bf q}\cdot {\bf r}}\psi_{\bf k}({\bf r})=(2\pi )^3\,\delta({\bf k-q})-
\frac{4\pi a_3({\bf q},{\bf k})}{q^2-k^2-i\delta}
\end{eqnarray}
Note that the second term has a sign opposite to the one found in \cite{stm}, just because our definition for $a_3$ has
an opposite sign. The scattering amplitude is obtained from $f_{k_0}(\theta)=-a_3({\bf q},{\bf k}_0)$ (with $|{\bf q}|=|{\bf k}_0|$).

When one takes the angular average of Eq.(\ref{wavef}), the resulting s-wave component $\psi_k^0(r)$ of the wave function
is given by:
\begin{eqnarray}\label{waves}
\psi_k^0(r)=\frac{e^{i\delta_0}}{kr}\sin(kr+\delta_0)
\end{eqnarray}
where $\delta_0(k)$ is the s-wave phase shift, linked to s-wave component $f_0(k)$ of the scattering amplitude by:
\begin{eqnarray}\label{scattamp}
f_0(k)=\frac{e^{2i\delta_0}-1}{2ik}
\end{eqnarray}

\subsection{Very light mass}\label{vlm}

Let us go now to the interesting regime where the mass of the lonely fermion $m\ds$ is very light (or equivalently
the mass $m\us$ of the two identical fermions is very heavy). In this case parameter $R$ is near zero and $R'$
is near $2$. In our preceding paper \cite{acl}, where we had ${\bar k}_0=0$, we have carefully expanded 
Eq.(\ref{eqa3bar}) near this limit. Then by a Fourier transform we have converted the resulting equation into a differential
equation with respect to a variable which is called ${\bar r}$ below. We have studied this last equation, matching in particular the behaviour for the small and the large
values of the variable ${\bar r}$ by appropriate consideration of the boundary conditions. However the important range, in order to obtain
the scattering length,
corresponds to the large ${\bar r}$ values and the proper behaviour in this range can also be obtained from physical considerations,
as we will see below. We can also proceed by continuity, requiring that in the present case the large variable $r$ behaviour reduces to
the one we have found in the limit ${\bar k}_0 \rightarrow 0$. More specifically we have seen in this preceding work that the careful expansion
of the right-hand side of Eq.(\ref{eqa3bar}) was useful only to satisfy the perfect matching with the boundary
conditions for the small values of the variable ${\bar r}$. On the other hand only the expansion of the left-hand side of
Eq.(\ref{eqa3bar}) was necessary in order to obtain the large ${\bar r}$ information necessary to obtain the
scattering length. Here we will take advantage of this fact to avoid the painfull task of expanding the right-hand side of Eq.(\ref{eqa3bar})
and rather rely on the above arguments to obtain the required large variable ${\bar r}$ behaviour. Hence we will set
$R$ and $R'$ to their limiting value, namely $R=0$ and $R'=2$ in the right-hand side. In the left-hand side,
following our preceding work, we will also set $1+ R\,({\bar k}^2-{\bar k}_0^2) \simeq 1$ because large values
of ${\bar k}^2-{\bar k}_0^2$ are irrelevant in our problem. However we will naturally keep the overall factor $R$
in front of the left-hand side. This is the only difference with taking the full $m\ds=0$ limit.

We rewrite Eq.(\ref{eqa3bar}) in this limit by making use, instead of $\bar{a}_3({\bar q})$, of the more
convenient function $F({\bar q})$ defined by:
\begin{eqnarray}\label{defF}
F({\bar q})=\frac{{\bar q}\,\bar{a}_3({\bar q})} {\bar{q}^2-\bar{k}_0^2-i\delta}
\end{eqnarray}
We note that, since from Eq.(\ref{eqa3bar}) $\bar{a}_3({\bar q})$ is an even function of ${\bar q}$,
$F({\bar q})$ is an odd function of ${\bar q}$. Then Eq.(\ref{eqa3bar}) becomes:
\begin{eqnarray}\label{eqa3leg}
\epsilon ^2(\bar{k}^2-\bar{k}_0^2)\,F({\bar k})
   &-&\frac{1}{4{\bar k}_0}\;\ln \frac{1+({\bar k}+{\bar k}_0)^2}
    {1+({\bar k}-{\bar k}_0)^2}
 \\ \nonumber
   &=-& \frac{1}{2\pi} \int_{0}^{\infty} d{\bar q}\;F({\bar q})\;\ln \frac{1+({\bar k}+{\bar q})^2}
    {1+({\bar k}-{\bar q})^2}
    =\frac{1}{2\pi} \int_{-\infty}^{\infty} d{\bar q}\;F({\bar q})\;\ln \left(1+({\bar k}-{\bar q})^2\right)
\end{eqnarray}
where in the last equality we have made use of the odd parity of $F({\bar q})$ and, 
as in \cite{acl}, we have set:
\begin{eqnarray}\label{defeps}
\epsilon ^2=\frac{R}{2} \simeq \frac{m\ds}{m\us}
\end{eqnarray}

We take now the Fourier transform of this equation, introducing:
\begin{eqnarray}\label{tFF}
{\bar F}({\bar r})= \int_{-\infty}^{\infty} d{\bar q}\, F({\bar q})\, \exp(-i{\bar q}{\bar r})
\end{eqnarray}
and making use of:
\begin{eqnarray}\label{tFln}
\int_{-\infty}^{\infty} d{\bar q}\, \ln \left(1+{\bar q}^2\right)\, \exp(-i{\bar q}{\bar r})=
-2\pi \frac{e^{-{\bar r}}}{\bar r}
\end{eqnarray}
where we have restricted ourselves to the case ${\bar r} > 0$.
We obtain:
\begin{eqnarray}\label{eqtF}
\epsilon ^2 \left(\frac{d^2}{d{\bar r}^2}+\bar{k}_0^2 \right){\bar F}({\bar r})=
\frac{e^{-{\bar r}}}{\bar r}\left({\bar F}({\bar r})+i\pi \frac{\sin(\bar{k}_0{\bar r})}{\bar{k}_0}\right)
\end{eqnarray}
Since the $\epsilon =0$ solution:
\begin{eqnarray}\label{F0}
{\bar F}_0({\bar r})=-i\pi \frac{\sin(\bar{k}_0{\bar r})}{\bar{k}_0}
\end{eqnarray}
satisfies $(d^2/d{\bar r}^2+{\bar k}_0^2){\bar F}_0({\bar r})=0$, it is convenient to set:
\begin{eqnarray}\label{defg}
{\bar F}({\bar r})={\bar F}_0({\bar r})+i\pi g({\bar r})
\end{eqnarray}
to obtain the simpler equation:
\begin{eqnarray}\label{eqg1}
\epsilon ^2\left(\frac{d^2}{d{\bar r}^2}+\bar{k}_0^2 \right)g({\bar r})=
\frac{e^{-{\bar r}}}{\bar r}g({\bar r})
\end{eqnarray}
The fact that we find a second order differential equation is directly related to
our lowest order expansion of Eq.(\ref{eqa3bar}) in $m\ds /m\us$. Going to higher
order in this expansion would result in a higher order differential equation

At this stage it is useful to notice the relation between $g({\bar r})$ and the scattering wavefunction
Eq.(\ref{waves}). Taking in Eq.(\ref{TF}) the angular average over the direction of ${\bf k}$, performing 
the angular integration over ${\bf r}$, setting 
$|{\bf k}|=|{\bf k}_0|$ and making use of the definition Eq.(\ref{defF}), we obtain:
\begin{eqnarray}\label{angav}
4\pi \int_{0}^{\infty}\!\!dr\, r \sin(qr)\psi_{k_0}^0(r)=\frac{2\pi ^2}{k_0}\left[\delta(q-k_0)-\delta(q+k_0)\right]
-4\pi F(q)
\end{eqnarray}
where we have added in the right-hand side the term $\delta(q+k_0)$ which is usely zero since
$q=|{\bf q}|>0$ and $k_0=|{\bf k}_0|>0$. However we have extended above the range of variation
of $q$ to negative values and made use of the fact that the corresponding extension of $F(q)$
is odd. The added term is necessary to satisfy in Eq.(\ref{angav}) this parity property of $F(q)$.
Going to reduced variables and taking the Fourier transform of Eq.(\ref{angav}) for ${\bar r}>0$
we obtain with Eq.(\ref{F0}):
\begin{eqnarray}\label{}
{\bar F}({\bar r})={\bar F}_0({\bar r})+
i\pi \frac{e^{i\delta_0}}{{\bar k}_0}\sin({\bar k}_0{\bar r}+\delta_0)
\end{eqnarray}
Hence from Eq.(\ref{defg}) we have for large ${\bar r}$ the very simple relation:
\begin{eqnarray}\label{gasymp}
g({\bar r})=\frac{e^{i\delta_0}}{{\bar k}_0}\sin({\bar k}_0{\bar r}+\delta_0)
\end{eqnarray}
This provides us with the boundary condition necessary to solve the second order
differential equation Eq.(\ref{eqg1}), which is accordingly an effective Schr\"odinger equation
for our scattering problem.

Except for the $\bar{k}_0^2$ term this equation Eq.(\ref{eqg1}) is the same as the one we have found in our
preceding work \cite{acl} and we solve it by following the same procedure. Performing the change of
variable $\bar{r}=2\ln(2/z)$, implying $z=2e^{-\bar{r}/2}$, we find that $\bar{g}(z) \equiv g(\bar{r}(z))$
satisfies:
\begin{eqnarray}\label{eqgz}
z^2\,\frac{d^2\bar{g}(z)}{dz^2}+z\,\frac{d\bar{g}(z)}{dz}+4\bar{k}_0^2\,\bar{g}(z)
-\frac{z^2}{L(z)}\, \bar{g}(z)=0
\end{eqnarray}
with $L(z)=2\epsilon ^2 \ln(2/z)=\epsilon ^2 \bar{r}$. We notice again that, in the range of variable which is of interest, that is
large $\bar{r}$, corresponding to small $z$, $L(z)$ is a quite slowly varying function of $z$. 
This allows to treat it as a constant $L(z) \simeq L$. As discussed in details in
\cite{acl} this approximate treatment is increasingly accurate when the mass ratio $m\ds / m\us$,
i.e. $\epsilon ^2$, becomes increasingly small, which is precisely the limit we are interested in. 
We have found \cite{acl} that the important range for matching the solution of Eq.(\ref{eqg1})
to the asymptotic behaviour, in the present case Eq.(\ref{gasymp}), is ${\bar r} \sim {\bar r}_0
=\ln (1/\epsilon ^2)$. Hence we may take $L=\epsilon ^2 \ln (1/\epsilon ^2)$.

Then the further change of variable $z=\sqrt{L}\,x$ transforms Eq.(\ref{eqgz}) into:
\begin{eqnarray}\label{eqgx}
x\,\frac{d^2\tilde{g}(x)}{dx^2}+\frac{d\tilde{g}(x)}{dx}+4\bar{k}_0^2\,\tilde{g}(x)-x \tilde{g}(x)=0
\end{eqnarray}
for $\tilde{g}(x)=\bar{g}(z)$.
The solution of this equation which matches, as we will see shortly, 
the asymptotic behaviour Eq.(\ref{gasymp}) is
\cite{gr} the well known Bessel functions $K_{2i\bar{k}_0}(x)$ (or equivalently $K_{-2i\bar{k}_0}(x)$),
which goes continuously to the solution $K_0(x)$ that we have found in \cite{acl} 
for the case $\bar{k}_0=0$. A convenient integral representation for this function is \cite{gr}:
\begin{eqnarray}\label{repr}
K_{2i\bar{k}_0}(x)= \int_{0}^{\infty}dt \,e^{-x\cosh t}\cos(2\bar{k}_0t)
\end{eqnarray}
From this representation one obtains the small $x$ behaviour:
\begin{eqnarray}\label{smalx}
K_{2i\bar{k}_0}(x) \simeq {\rm Re}\left[\left(\frac{2}{x}\right)^{2i\bar{k}_0}\Gamma(2i\bar{k}_0) \right]
\end{eqnarray}
where $\Gamma(x)$ is the Euler function. From this result one recovers in particular, 
for $\bar{k}_0 \to 0$, the
known small $x$ asymptotic behaviour $K_0(x) \simeq \ln(2/x)-C$ where $C$ is the Euler constant.

Since $2/x=\sqrt{L}\,e^{{\bar r}/2}$ we see that, for large ${\bar r}$, our solution has indeed the required
asymptotic behaviour Eq.(\ref{gasymp}). 
This is naturally within an unimportant constant prefactor since the homogeneous
differential equation Eq.(\ref{eqgx}) does not allow to find this prefactor.
But, as soon as $\delta_0$ is found, this prefactor is given explicitly by Eq.(\ref{gasymp}).
Comparing Eq.(\ref{smalx}) with Eq.(\ref{gasymp}) in order to identify the phase shift, we find easily:
\begin{eqnarray}\label{delta0}
\delta_0(\bar{k}_0)=\bar{k}_0 \ln L+{\rm Arg}\left[i\Gamma(2i\bar{k}_0)\right]=\bar{k}_0
\ln\left[\epsilon ^2 \ln (1/\epsilon ^2)\right]+{\rm Arg}\left[i\Gamma(2i\bar{k}_0)\right]
\end{eqnarray}
In particular in the limit $\bar{k}_0 \rightarrow 0$ we can make use of $\Gamma(2i\bar{k}_0)
\simeq 1/(2i\bar{k}_0)-C$, implying ${\rm Arg}\left[i\Gamma(2i\bar{k}_0)\right] \simeq -2\bar{k}_0C$. 
Comparing to the limiting behaviour $\delta_0(\bar{k}_0) \simeq
-\bar{a}_3\bar{k}_0$ we recover our result \cite{acl}:
\begin{eqnarray}\label{a3}
\bar{a}_3 \equiv \frac{a_3}{a}=-\ln L+2C=\ln\frac{m\us}{m\ds}-\ln \ln\frac{m\us}{m\ds}+2C
\end{eqnarray}
for the fermion-dimer scattering length in the considered limit where the mass ratio 
$m\us / m\ds$ is large.

Once the phase shift is known we obtain the scattering amplitude by merely making use
of Eq.(\ref{scattamp}). This is obviously the simpler way. However let us briefly show that
we recover this result by calculating directly $a_3({\bf q},{\bf k}_0)$ from our solution 
$K_{2i\bar{k}_0}(x)$ of Eq.(\ref{eqgx}). From section \ref{base} we have:
\begin{eqnarray}\label{f0}
{\bar f}_0(k_0) \equiv \frac{f_0(k_0)}{a}=-{\rm lim}_{{\bar q} \to {\bar k}_0}\,\bar{a}_3({\bar q})
=-2\,{\rm lim}_{{\bar q} \to {\bar k}_0}\,\left({\bar q}-{\bar k}_0\right)F({\bar q})
\end{eqnarray}
the last equality resulting from Eq.(\ref{defF}). In the Fourier transform of the decomposition Eq.(\ref{defg})
we see that $F_0({\bar q})$ does not contribute to the right-hand side of Eq.(\ref{f0})
since, as we have seen in Eq.(\ref{angav}):
\begin{eqnarray}\label{}
F_0({\bar q})=\frac{\pi}{2{\bar k}_0}\left[\delta({\bar q}-{\bar k}_0)-\delta({\bar q}+{\bar k}_0)\right]
\end{eqnarray}
so that, in this right-hand side, we are left only with the contribution of $g$:
\begin{eqnarray}\label{f0bis}
{\bar f}_0(k_0)
=-2\,{\rm lim}_{{\bar q} \to {\bar k}_0}\,\left({\bar q}-{\bar k}_0\right)
i\pi \int_{-\infty}^{\infty}\frac{d{\bar r}}{2\pi }\,g({\bar r})\,\exp(i{\bar q}{\bar r})
\end{eqnarray}

\begin{figure}
\centering
{\includegraphics[width=\linewidth]{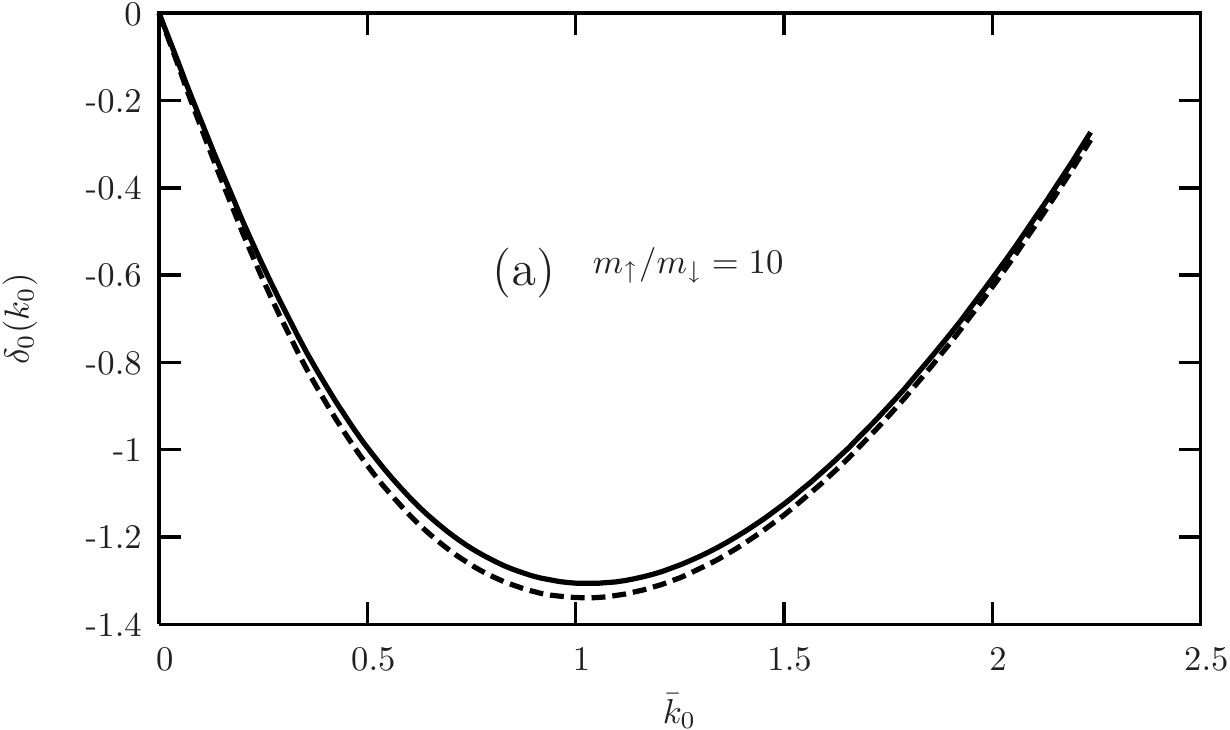}}
{\includegraphics[width=\linewidth]{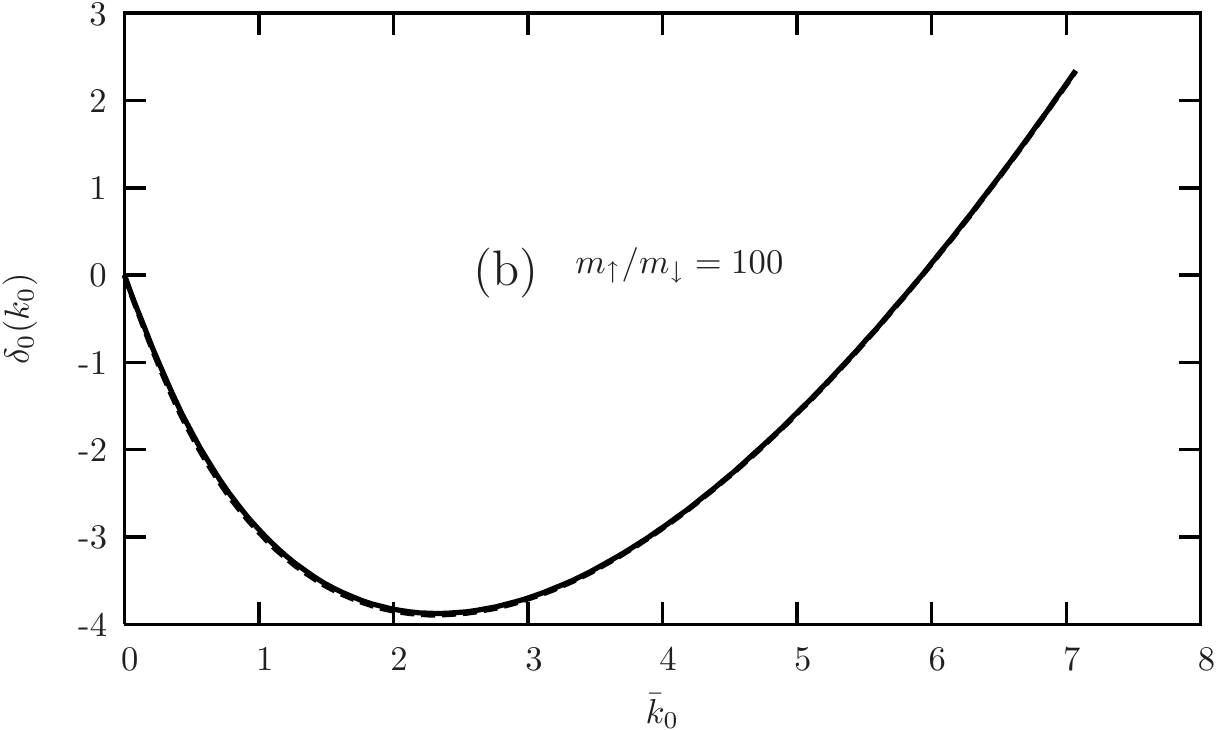}}
\caption{$\ell=0$ phase shift $\delta_0(k_0)$ as a function of $\bar{k}_0=k_0\,a$, given by Eq.(\ref{delta0}) (dashed
line) and its asymptotic approximation given by Eq.(\ref{delta01}) (full line) for a) $m\us /m\ds = 10$ and
b)  $m\us /m\ds =100$.}
\label{fig1s}
\end{figure}

We have seen that the solution of Eq.(\ref{eqgx}) is proportional to $K_{2i\bar{k}_0}(x)$. Going back to the 
${\bar r}$ variable this translates into:
\begin{eqnarray}\label{eqgr}
g({\bar r})=\alpha ({\bar k}_0) K_{2i\bar{k}_0}\left(\frac{2e^{-\bar{r}/2}}{\sqrt{L}}\right)
\end{eqnarray}
where, as mentionned above, the prefactor $\alpha ({\bar k}_0)$ can not be obtained from the homogeneous equation Eq.(\ref{eqgx})
and has to be determined from the asymptotic behaviour Eq.(\ref{gasymp}) for $g({\bar r})$. From the asymptotic behaviour Eq.(\ref{smalx})
we have:
\begin{eqnarray}\label{}
K_{2i\bar{k}_0}\left(\frac{2e^{-\bar{r}/2}}{\sqrt{L}}\right) \simeq |\Gamma(2i\bar{k}_0)|\,\sin({\bar k}_0{\bar r}+\delta_0)
\end{eqnarray}
where $\delta_0$ is given by Eq.(\ref{delta0}). Comparing Eq.(\ref{eqgr}) with Eq.(\ref{gasymp}) we find:
\begin{eqnarray}\label{alpha}
\alpha ({\bar k}_0)=e^{i\delta_0}\frac{1}{{\bar k}_0|\Gamma(2i\bar{k}_0)|}
\end{eqnarray}

The Fourier transform of $g({\bar r})$ is then:
\begin{eqnarray}\label{}
 \int_{-\infty}^{\infty}\frac{d{\bar r}}{2\pi }\,g({\bar r})\,\exp(i{\bar q}{\bar r})= \int_{0}^{\infty}\frac{d{\bar r}}{2\pi }\,g({\bar r})\,\exp(i{\bar q}{\bar r})
 -\left({\bar q} \rightarrow -{\bar q} \right)
\end{eqnarray}
where we have made use of the odd parity of $g({\bar r})$ directly linked to the odd parity of $F({\bar q})$.
Going back to the variable $x=2e^{-\bar{r}/2}/\sqrt{L}$ we have:
\begin{eqnarray}\label{}
 \int_{0}^{\infty}\frac{d{\bar r}}{2\pi }\,g({\bar r})\,\exp(i{\bar q}{\bar r})&=&\frac{\alpha ({\bar k}_0)}{2\pi } \int_{0}^{\infty}d{\bar r}\,
\exp(i{\bar q}{\bar r}) \,K_{2i\bar{k}_0}\left(\frac{2e^{-\bar{r}/2}}{\sqrt{L}}\right) \\ \nonumber
&=&\frac{\alpha ({\bar k}_0)}{\pi }\,\left(\frac{\sqrt{L}}{2}\right)^{-2i{\bar q}}\int_{0}^{\infty}dx\,x^{-2i{\bar q}-1}K_{2i\bar{k}_0}(x)
\end{eqnarray}
where in the last step we have replaced the upper bound $2/\sqrt{L}$ by $\infty$, since in our case $L=\epsilon ^2 \ln (1/\epsilon ^2) \ll 1$
and, as it is clear from Eq.(\ref{repr}), $K_{2i\bar{k}_0}(x)$ decreases exponentially rapidly for large $x$. The last integral can be performed
analytically \cite{gr}, leading to:
\begin{eqnarray}\label{tffin}
 \int_{-\infty}^{\infty}\frac{d{\bar r}}{2\pi }\,g({\bar r})\,\exp(i{\bar q}{\bar r})=\frac{\alpha ({\bar k}_0)}{4\pi }
 L^{-i{\bar q}}\Gamma\left(i({\bar k}_0-{\bar q})\right)\Gamma\left(-i({\bar k}_0+{\bar q})\right) -\left({\bar q} \rightarrow -{\bar q} \right)
\end{eqnarray}
Since $\Gamma(x)\simeq 1/x$ for $x \to 0$, we find from Eq.(\ref{f0bis}) and Eq.(\ref{tffin}):
\begin{eqnarray}\label{}
{\bar f}_0(k_0)=\frac{\alpha ({\bar k}_0)}{2}\left[L^{i{\bar k}_0}\Gamma(2i{\bar k}_0)+
L^{-i{\bar k}_0}\Gamma(-2i{\bar k}_0)\right]
\end{eqnarray}
However we have from Eq.(\ref{delta0}):
\begin{eqnarray}\label{}
\sin(\delta_0(\bar{k}_0))=L^{i{\bar k}_0}\frac{\Gamma(2i{\bar k}_0)}{2|\Gamma(2i\bar{k}_0)|}+
L^{-i{\bar k}_0}\frac{\Gamma(-2i{\bar k}_0)}{2|\Gamma(-2i\bar{k}_0)|}
\end{eqnarray}
Taking Eq.(\ref{alpha}) into account this leads to:
\begin{eqnarray}\label{}
{\bar f}_0(k_0)=e^{i\delta_0}\frac{\sin(\delta_0(\bar{k}_0))}{{\bar k}_0}
\end{eqnarray}
which, when we go back to unreduced units, is identical to Eq.(\ref{scattamp}) as it should be.
This shows that our calculations and approximations in the very light mass limit are fully consistent.

\subsection{Discussion}\label{disc}

Let us now consider the implications of our result Eq.(\ref{delta0}) for the phase shift and the corresponding scattering amplitude.
Although the Euler $\Gamma$ function is easily obtained numerically, we have more insight if we make use of an excellent
approximation. Since for real $x$ we have:
\begin{eqnarray}\label{}
{\rm Arg}\left[i\Gamma(ix)\right]={\rm Arg}\left[ix\Gamma(ix)\right]={\rm Arg}\left[\Gamma(1+ix)\right]
={\rm Im}\left[\ln\Gamma(1+ix)\right]
\end{eqnarray}
we may use the excellent Stirling formula $\ln \Gamma(z) \simeq (z-1/2)\ln z -z+(1/2) \ln(2\pi )$ which leads to:
\begin{eqnarray}\label{delta01}
\delta_0(\bar{k}_0)=-\bar{k}_0 \bar{a}_3+\frac{1}{2} \arctan(2\bar{k}_0)+2(C-1)\bar{k}_0+\bar{k}_0 \ln(1+4\bar{k}_0^2)
\end{eqnarray}
If we were to correct the small imperfection of the Stirling formula for low $z$ by replacing $2C \simeq 1.154$ by $1$,
this would allow Eq.(\ref{delta01}) to give the exact result ${\rm lim}_{\bar{k}_0 \to 0}\delta_0(\bar{k}_0)=-\bar{k}_0 \bar{a}_3$
with $\bar{a}_3$ given by Eq.(\ref{a3}), but the result would not be accurate for higher values of $\bar{k}_0$. 
Our original result Eq.(\ref{delta0}) and its simplified form Eq.(\ref{delta01}) are
displayed in Fig.\ref{fig1s} for $m\us /m\ds = 10$ and $100$. In this last case the two results are essentially undistinguishable
in the figure.

\begin{figure}
\centering
{\includegraphics[width=\linewidth]{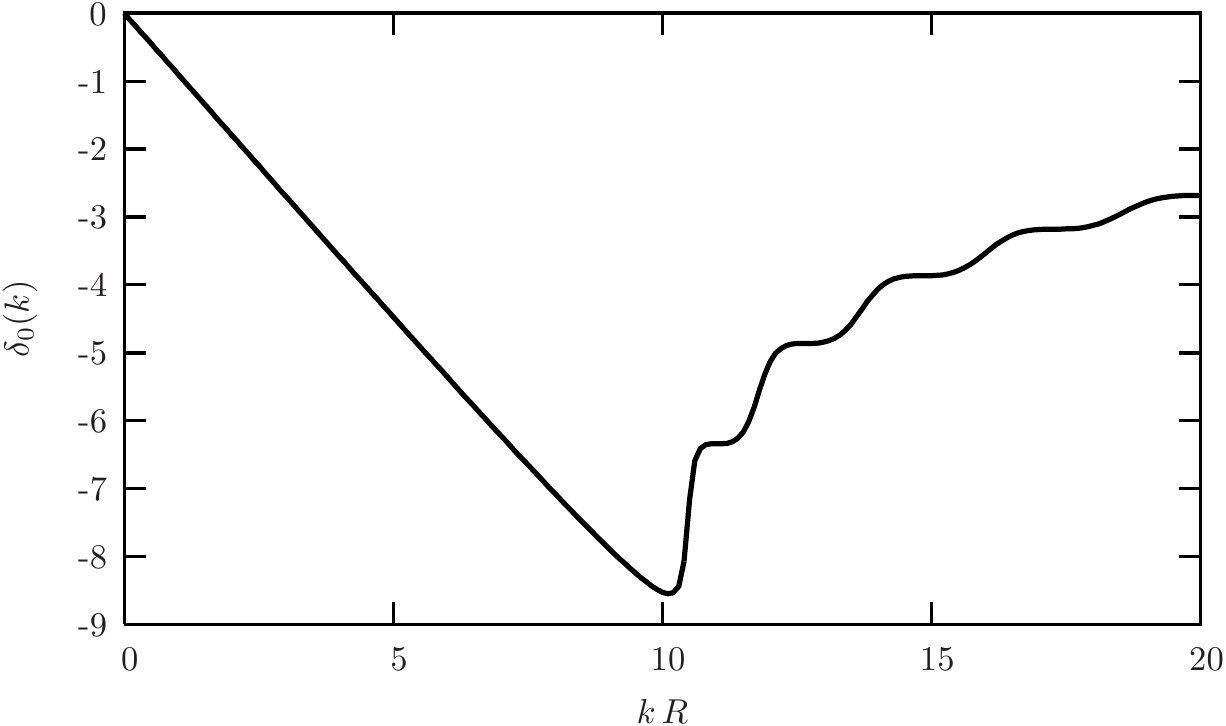}}
\caption{$\ell=0$ phase shift $\delta_0(k)$ for a repulsive square potential with strength
$V_0$ and range $R$, as a function of $kR$, in the case $(2mV_0)^{1/2}R/\hbar=10$. The oscillations beyond the maximum are due to the discontinuous
rise of the potential.}
\label{fig2s}
\end{figure}

We can see that $\delta_0(\bar{k}_0)$ displays first a strong decrease from zero down to large negative values after which
it increases back to small negative values, and even becomes positive in the case of large mass ratios. For large $\bar{a}_3$
approximate analytical results may be obtained for the location of
the minimum and its value. They both scale as $\sqrt{m\us /m\ds}$. The essential qualitative feature of this result,
namely the strong decrease of the phase shift, is directly linked to the large positive value of the scattering length.
For example it is qualitatively fairly analogous to the one found for the $\ell=0$ phase shift of a square repulsive potential with strength
$V_0$ and range $R$, which is given by:
\begin{eqnarray}\label{delta01sq}
\delta_0(k)=-kR+ \arctan\left[\frac{k}{k'}\tan(k'R)\right]
\end{eqnarray}
where $k$ is related to the energy $E$ by $E=\hbar^2 k^2/2m$ and $E-V_0=\hbar^2 k'^2/2m$. This is plotted in Fig.\ref
{fig2s} for $(2mV_0)^{1/2}R/\hbar=10$.

\begin{figure}
\centering
{\includegraphics[width=\linewidth]{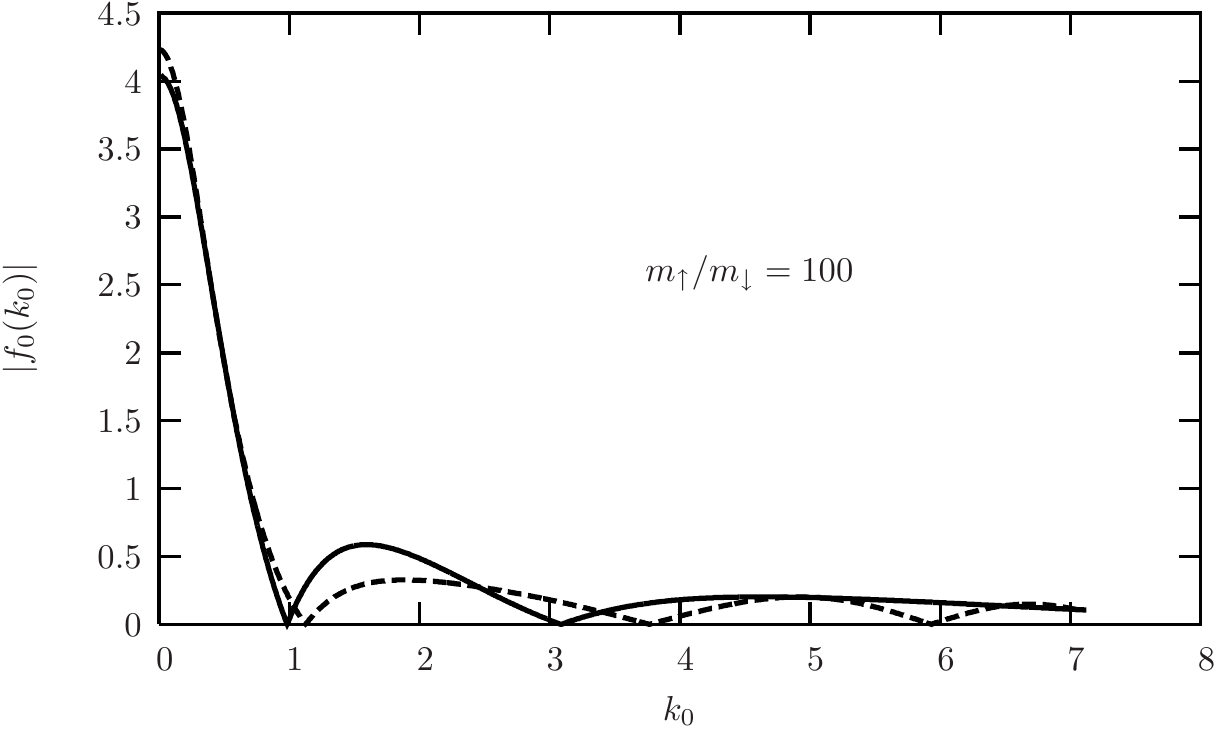}}
\caption{Modulus of the $\ell=0$ scattering amplitude $|f_0(k_0)|$ as a function of $k_0$, for mass ratio
$m\us / m\ds=100$. Dashed line: asymptotic result. Full line: exact numerical result.}
\label{fig3s}
\end{figure}

\begin{figure}
\centering
{\includegraphics[width=\linewidth]{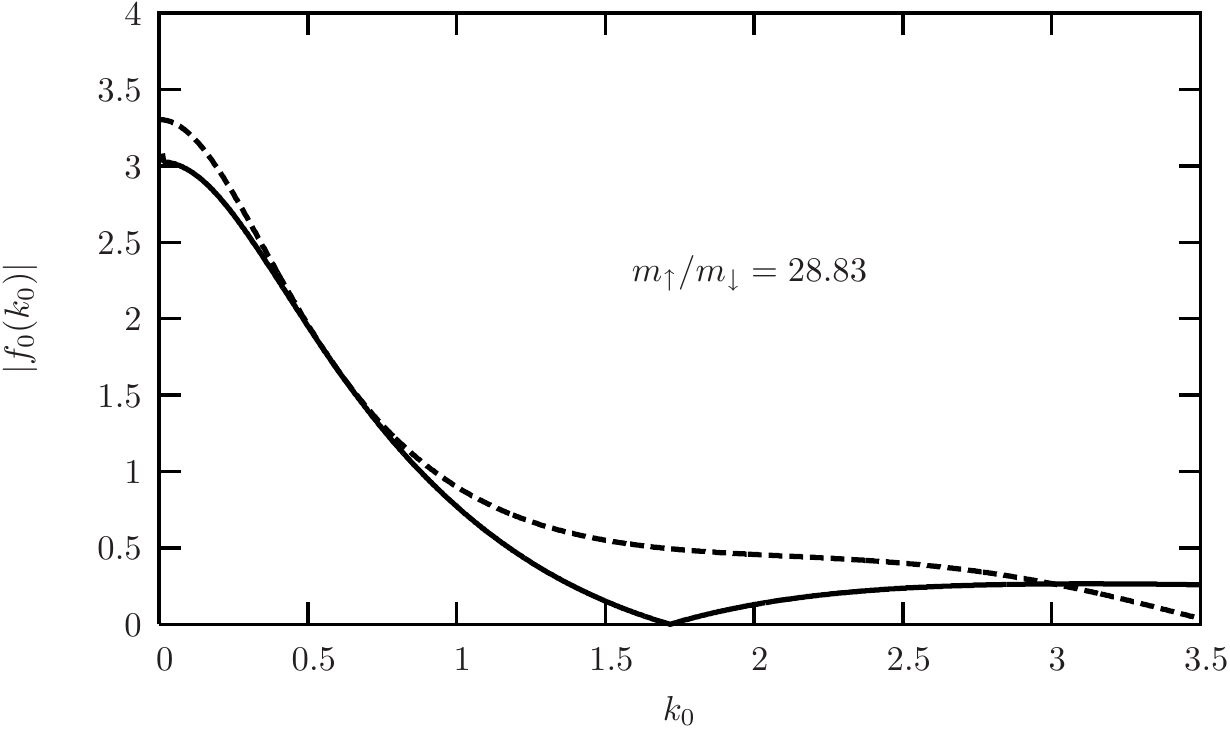}}
\caption{Modulus of the $\ell=0$ scattering amplitude $|f_0(k_0)|$ as a function of $k_0$, for mass ratio
$m\us / m\ds=28.83$ corresponding to a hypothetical $^6$Li - $^{173}$Yb mixture. Dashed line: asymptotic result. Full line: exact numerical result.}
\label{fig4s}
\end{figure}

\begin{figure}
\centering
{\includegraphics[width=\linewidth]{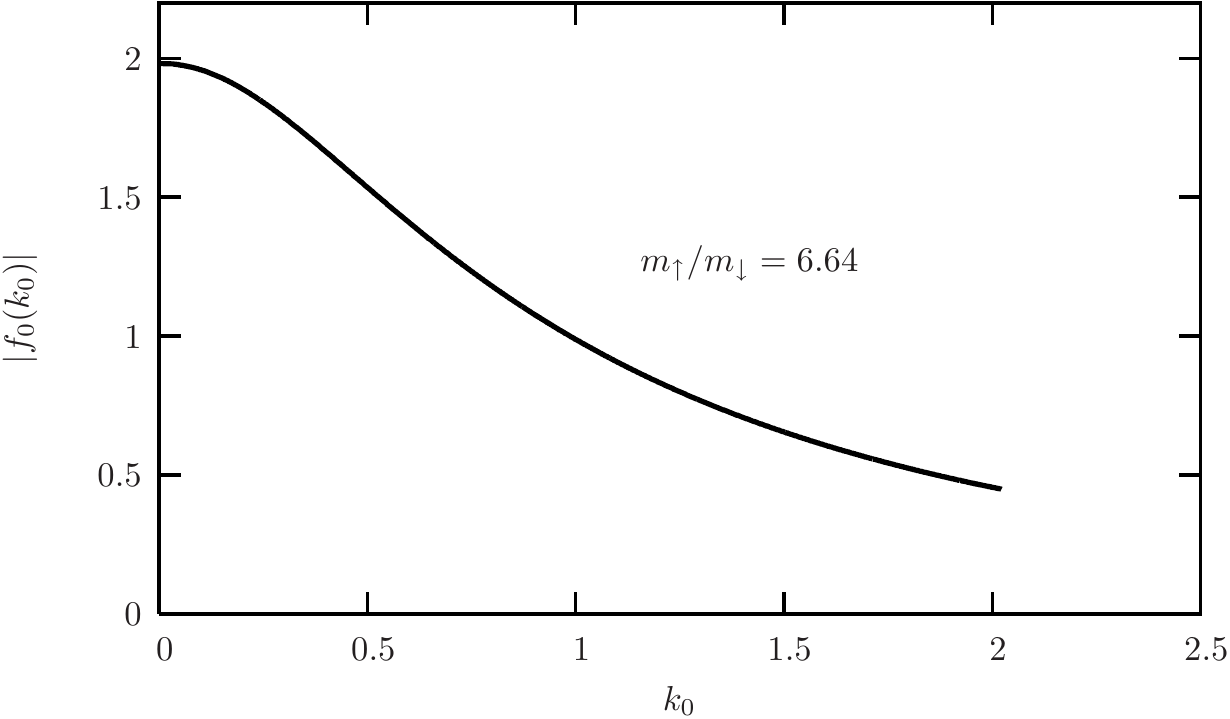}}
\caption{Modulus of the $\ell=0$ scattering amplitude $|f_0(k_0)|$ as a function of $k_0$, for mass ratios
$m\us / m\ds=6.64$ corresponding to a $^6$Li - $^{40}$K mixture.}
\label{fig5s}
\end{figure}

We consider now the consequences of the above result for the s-wave scattering amplitude which is obtained from Eq.(\ref{scattamp}).
We are mostly interested in its modulus $|{\bar f}_0(k_0)|=|\sin(\delta_0(\bar{k}_0))|/\bar{k}_0$. The result is plotted in Fig.~\ref{fig3s} for
$m\us /m\ds = 100$. The very interesting feature is the existence of several zeros occuring when $\delta_0(\bar{k}_0)=n\pi $.
The first few zeros are merely located at $\bar{k}_0=n\pi /\bar{a}_3$.
Physically these zeros imply that there is no contribution
from s-wave scattering to the total cross section. We see also that the scattering amplitude is strongly peaked for zero energy, a feature which
is even more important for the corresponding contribution $\sigma_0(k_0)=4\pi a^2 {\bar f}^2_0(\bar{k}_0)$ to the cross section.
This feature is clearly linked to the strong decrease of the phase shift when $\bar{k}_0$ starts from zero.
As mentionned above this is directly linked to the large value of the scattering length, and consequently this is clearly not an artefact
of our approximate treatment. This behaviour implies that a rough model for the s-wave component of the effective fermion-dimer
interaction could be a constant up to the energy $(\hbar \pi /2a_3)^2/2\mu _T$ followed by no interaction at all for higher relative kinetic
energy. This is quite different from the simple picture of a constant interaction, whatever the energy, which is valid for example
when the fermions have equal masses.

These results we have obtained appear quite naturally in the small $m\ds / m\us$ limit. Their qualitative features are interesting
enough to check that they are not just a curiosity of this simple, but not so physical, limit and that they rather stay valid qualitatively
in the physical range for atomic mass ratios.
We have checked that our result is in semiquantitative agreement with the exact numerical result from Eq.(\ref{eqa3bar}). This is displayed
first for $m\us /m\ds = 100$ which is somewhat beyond realistic values. We see from Fig.~\ref{fig3s} that our approximation is in fair
agreement with the exact one for reasonably small energies (the discrepancy at and near zero energy is just due to the difference between
our approximate value for the scattering length and the exact one). There is however an increasing disagreement for higher values of the
energy, which is not so surprising. Our approximate phase shift is varying with energy somewhat less rapidly than the exact one.

If we go now to the more realistic value corresponding to a hypothetical mixture of $^6$Li and $^{173}$Yb displaying a Feshbach
resonance, we obtain for this mass ratio $m\us /m\ds = 173/6 \simeq 29$ the results reported in Fig.~\ref{fig4s}. It is quite interesting
to see that the results are not deeply modified compared to the low energy part of the preceding figure. Both the approximate and
the exact result display a single zero. Just as in the preceding case, the exact zero is at a somewhat lower energy than what our
approximate analytical expression gives. It is interesting that the relative location of this zero is in this order. Indeed our figure
shows that the approximate zero is near the dissociation threshold of the dimer, and that it is no longer present for slightly lower
values of the mass ratio $m\us /m\ds$. On the other hand the exact zero is only halfway to the dissociation threshold. 

Since the
existence of this zero is a landmark of a strongly varying s-wave scattering amplitude, it is of high interest that it exists for even
lower, and accordingly more realistic, mass ratio $m\us /m\ds$. We have found specifically that the exact zero is disappearing for
$m\us /m\ds \simeq 15.27$. If we take the lighter fermion to be $^6$Li, this would imply for the heavier one a mass quite near
the one of Rubidium. Unfortunately the two stable isotopes of Rubidium are bosonic, but there are two fermionic ones, $^{86}$Rb
and $^{84}$Rb, which have half-life of order of a month. Finally if we take the case of the 
$^6$Li - $^{40}$K mixture, which is displayed in Fig.~\ref{fig5s}, we see that the scattering cross section (obtained by squaring the
amplitude) near the dissociation threshold is only 5\% of its zero energy value.

Finally it is worth noting that the disappearance of this zero in the scattering amplitude when the mass ratio is lowered is strikingly
similar to the same disappearance \cite{acl,isk} in the function $a_3({\bf k},{\bf 0})$, which comes in the calculation of the scattering length
$a_3$. In this last case the disappearance occurs at a somewhat lower value of the mass ratio, and $a_3({\bf k},{\bf 0})$ has no simple
physical meaning, which makes the interpretation difficult. However one could go continuously from this function to the scattering amplitude
by varying the last parameter from ${\bf 0}$ to ${\bf k}$. Hence the two functions are closely related, and the simple physical explanation for the
zero in the scattering amplitude (it occurs when the phase shift reaches $\pi $) gives a corresponding understanding for $a_3({\bf k},{\bf 0})$.

\begin{figure}
\centering
{\includegraphics[width=\linewidth]{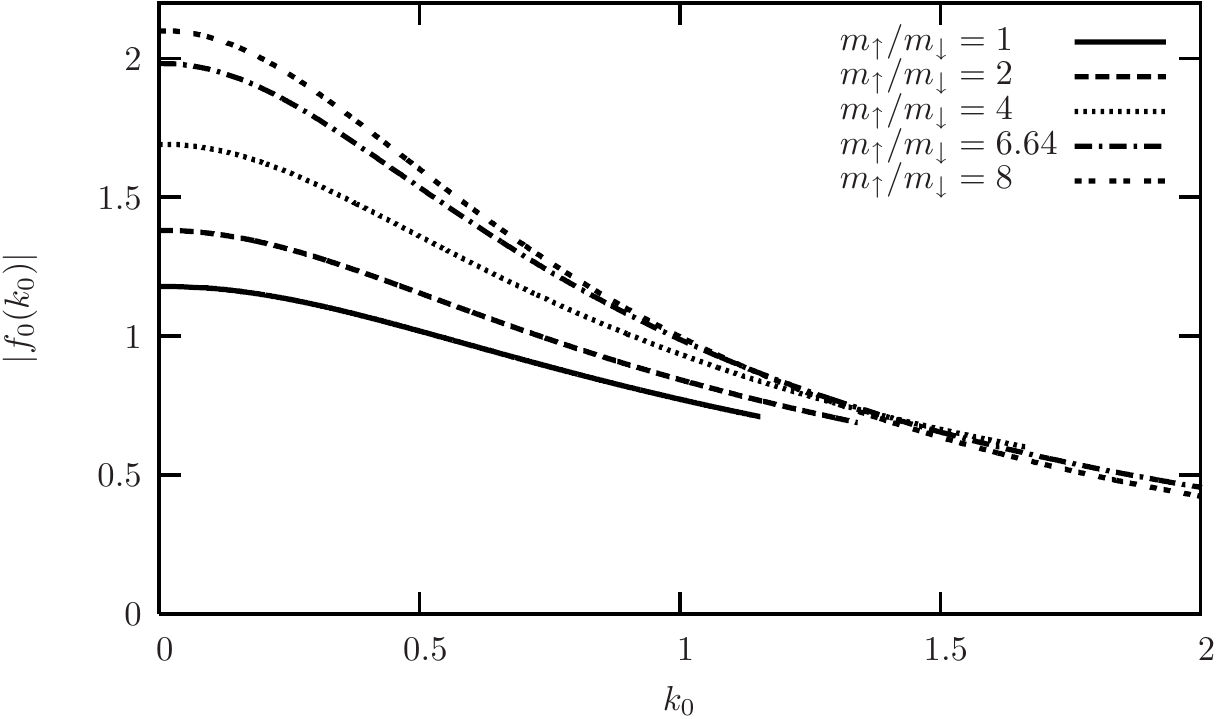}}
\caption{Modulus of the $\ell=0$ scattering amplitude $|f_0(k_0)|$ as a function of $k_0$, for mass ratios
$m\us / m\ds=1, 2, 4, 6.64$ and $8$. The ratio $m\us / m\ds=6.64$ corresponds to the $^6$Li - $^{40}$K
mixture. All these quantities are in reduced units, equivalent to set $a=1$.}
\label{fig1}
\end{figure}

\begin{figure}
\centering
{\includegraphics[width=\linewidth]{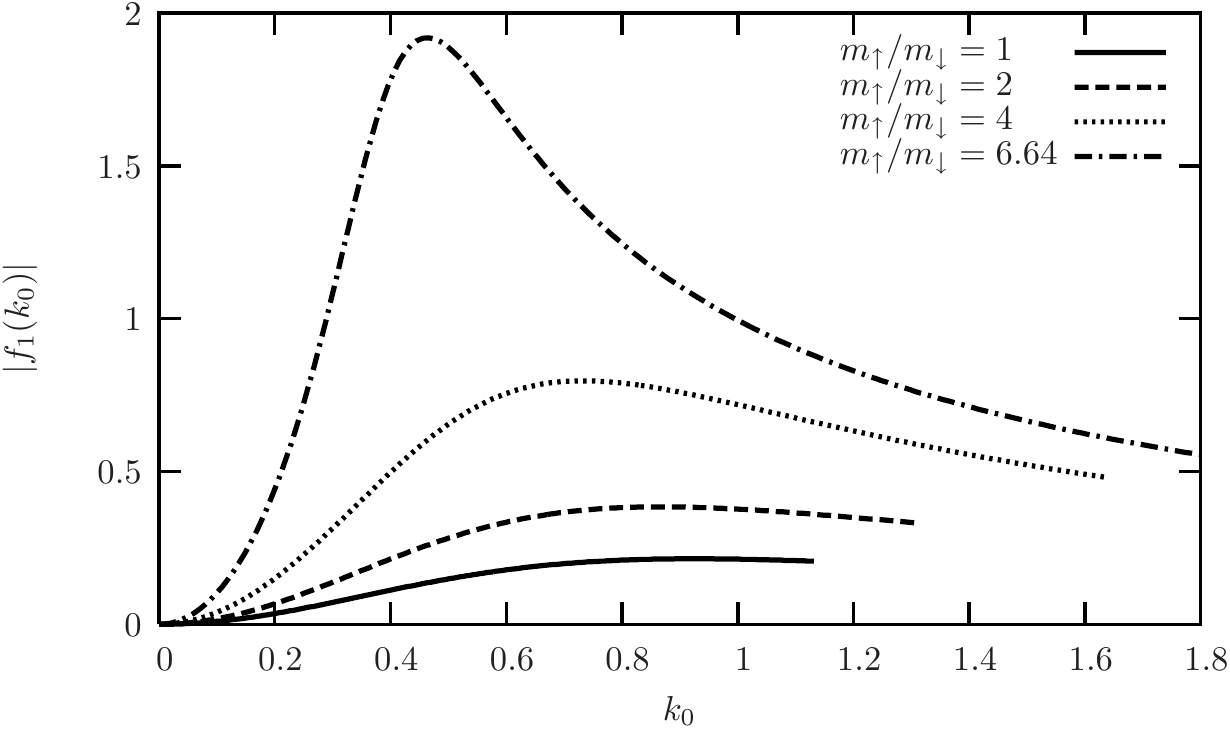}}
\caption{Modulus of the $\ell=1$ scattering amplitude $|f_1(k_0)|$ as a function of $k_0$, for mass ratios
$m\us / m\ds=1, 2, 4$ and $6.64$. The ratio $m\us / m\ds=6.64$ corresponds to the $^6$Li - $^{40}$K mixture.}
\label{fig2}
\end{figure}

\section{p - wave scattering}\label{pwav}

We turn now to the p-wave contribution to the scattering amplitude.
The analysis is done with the Skorniakov and Ter-Martirosian equations as detailed in section \ref{base}.
The only difference is that, instead of averaging Eq.(\ref{defa3}) to obtain the s-wave component,  in order to obtain the p-wave component
$a_{3 \rm p}(k,k_0)=(1/4\pi ) \int d\Omega_{\bf k} P_1({\bf k},{\bf k}_0) a_{3}({\bf k},{\bf k}_0)$, where
$P_1({\hat {\bf k}},{\hat {\bf k}}_0)={{\bf k}} \cdot {\bf k}_0/kk_0$ is the $\ell=1$ Legendre polynomial,
we have to project on this $\ell=1$ Legendre polynomial.

The resulting equation is quite similar to the s-wave one Eq.(\ref{eqa3bar}).
This is seen quite clearly by introducing the notations (making again use of reduced variables,
but omitting systematically
the bar over quantities in reduced units):
\begin{eqnarray}\label{}
\alpha (k,q)&=&1-Rk_0^2+k^2+q^2 \\ \nonumber
\beta (k,q)&=&R'kq
\end{eqnarray}
and defining the kernels:
\begin{eqnarray}\label{}
K_{\rm s}(k,q)=\frac{1}{2\beta(k,q)}\ln \left|\frac{\alpha (k,q)+\beta (k,q)}{\alpha (k,q)-\beta (k,q)}\right|
\end{eqnarray}
and
\begin{eqnarray}\label{}
K_{\rm p}(k,q)=\frac{1}{\beta (k,q)}-\frac{\alpha (k,q)}{2\beta^2 (k,q)}\ln \left|\frac{\alpha (k,q)+\beta (k,q)}{\alpha (k,q)-\beta (k,q)}\right|
\end{eqnarray}
Then the equations for the s-wave and p-wave components can be written:
\begin{eqnarray}\label{inteq}
\!\!R\;\frac{a_{3 \rm s,p}(k,k_0)}{1+\sqrt{1+ R\,(k^2-k_0^2)}}\!\!&=&\!\!K_{\rm s,p}(k,k_0) \nonumber \\
- \frac{2}{\pi} \int_{0}^{\infty} dq\!\!&&\!\!\frac{q^2 a_{3 \rm s,p}(q,k_0)}{q^2-k_0^2-i\delta}\;K_{\rm s,p}(k,q)
\end{eqnarray}
where $a_{3 \rm s}(k,k_0) \equiv \bar{a}_3({\bar k})$ used in section \ref{swave}.
 
Since we are interested in this part in mass ratios smaller than in the preceding section \ref{swave},
we provide first the results for the s-wave component for the specific mass ratios of interest. 
The results for $|f_0(k_0)|$ in terms of $k_0$ are given in Fig.\ref{fig1}, for the mass ratios $m\us / m\ds=
1, 2, 4, 8$. We have also inserted the result for $m\us / m\ds=6.64$ corresponding to the $^6$Li - $^{40}$K
mixture, already displayed in Fig.~\ref{fig5s}.
Except for the fairly slow increase of the atom-dimer scattering length with the mass ratio found at $k_0=0$, 
all the results are of order unity for $k_0=1$ and keep decreasing for increasing $k_0$.

\begin{figure}
\centering																	
{\includegraphics[width=\linewidth]{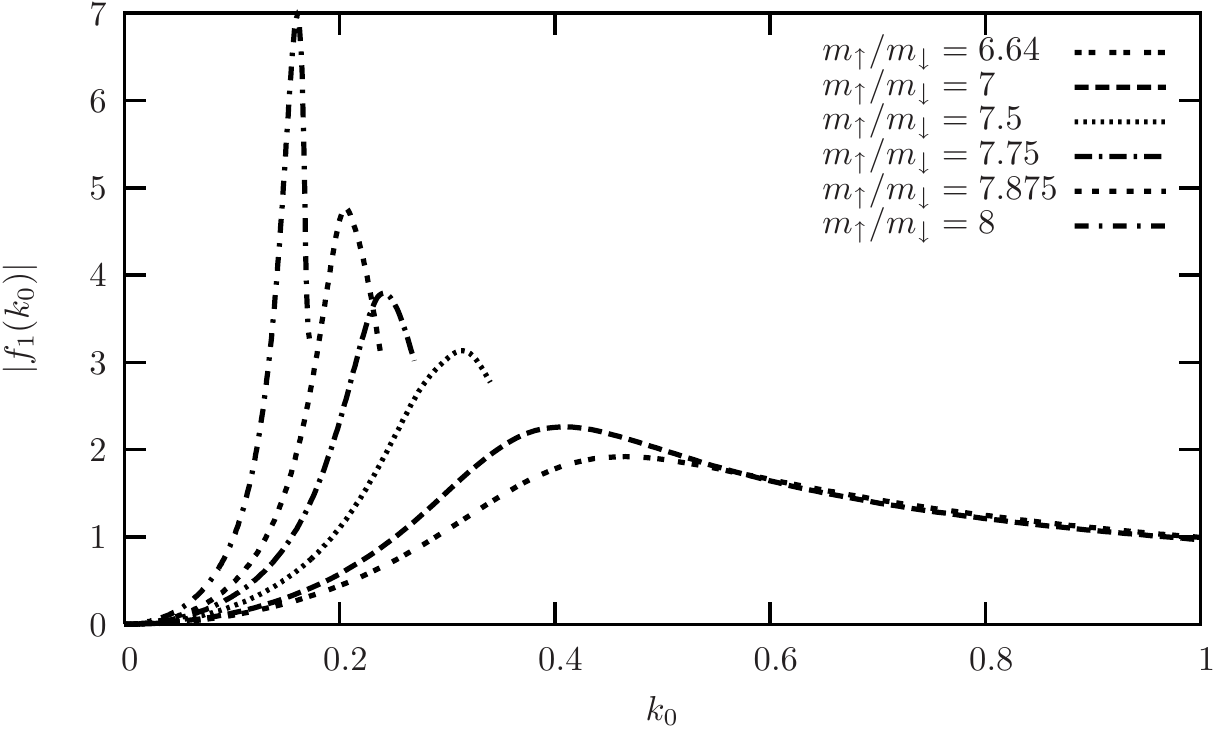}}
\caption{Modulus of the $\ell=1$ scattering amplitude $|f_1(k_0)|$ as a function of $k_0$, for mass ratios
$m\us / m\ds=6.64$ (corresponding to the $^6$Li - $^{40}$K mixture), $7, 7.5, 7.75, 7.875$ and $8$.}
\label{fig3}
\end{figure}

We turn now to the p-wave component results. We first give in Fig.\ref{fig2} the results for mass ratios
$m\us / m\ds=1, 2, 4$ and $6.64$, stopping at the value corresponding to the $^6$Li - $^{40}$K
mixture. For equal masses the p-wave component is always small, as it could be
expected, and it can clearly be omitted when
one deals with the atom-dimer scattering properties. Hence the atom-dimer vertex can, to a large extent,
be taken as a constant proportional to the atom-dimer scattering length. No complication is thus expected
to arise from this side when many-body properties will be investigated.  The situation is qualitatively similar for $m\us / m\ds= 2$ 
and even $4$: the p-wave component is fairly small and can be neglected compared to the s-wave contribution. 
However, we see that this p-wave
component rises steeply when the mass ratio goes from $4$ to $6.64$. This last ratio is at the border of
the resonant domain. 

\begin{figure}
\centering
{\includegraphics[width=\linewidth]{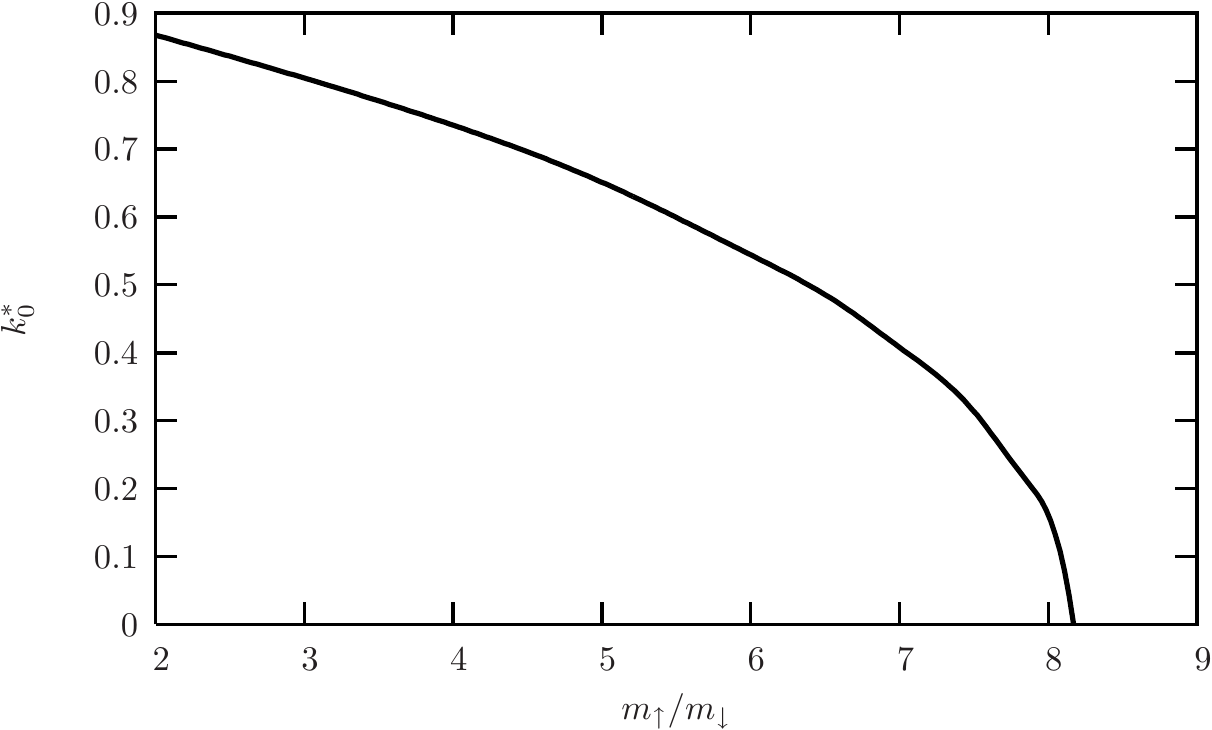}}
\caption{Location of the maximum $k_0^*$ of $|f_1(k_0)|$ as a function of mass ratio $m\us / m\ds$.}
\label{fig4}
\end{figure}

This domain is now displayed in Fig.~\ref{fig3},  where the results for mass ratios
$m\us / m\ds=6.64, 7., 7.5, 7.75, 7.875$ and $8.$ are plotted.
We see that the hump, present for $m\us / m\ds=6.64$ around $k_0 \simeq 0.45$, develops into a strong resonance
at lower and lower energy when $m\us / m\ds$ is increased. Although we have sampled mass ratios which are very
close to each other, we see that the resonance grows very rapidly with increasing mass ratio. The physical origin
is clearly the development of a virtual bound state at positive energy. The lifetime of this state is directly related to the width
of the resonance. Since the resonance peak gets quickly narrower, the lifetime grows very rapidly with increasing mass.
The link with the bound state which appears \cite{karma} at zero energy for $m\us / m\ds=8.17$ is confirmed if we
look at the position of the resonance peak as a function of the mass ratio. This is shown in Fig.\ref{fig4}.
We see that the position $k_0^*$ of the resonance peak extrapolates to zero for a mass ratio which is in close vicinity of $8.17$.
More precisely when the resonance peak reaches $k_0^*=0$, it will be infinitely sharp, corresponding to an infinite value
for $|f_1(k_0=0)|$. Hence in this case $a_{3 \rm p}(k,0)$ will be infinitely large, which implies that the homogeneous
Eq.(\ref{inteq}) (i.e. without the term $K_{\rm p}(k,0)$ in the right-hand side) has a solution for zero energy $k_0=0$.
This is just stating the well-known result that bound states are solutions of the homogeneous integral equation
corresponding to Eq.(\ref{inteq}). It is easy to find the lowest $m\us / m\ds$ for which this homogeneous integral equation
has a solution. We find that this occurs for $m\us / m\ds=8.172$ in full agreement with Kartavtsev and Malykh \cite{karma}.

\begin{figure}
\centering
{\includegraphics[width=\linewidth]{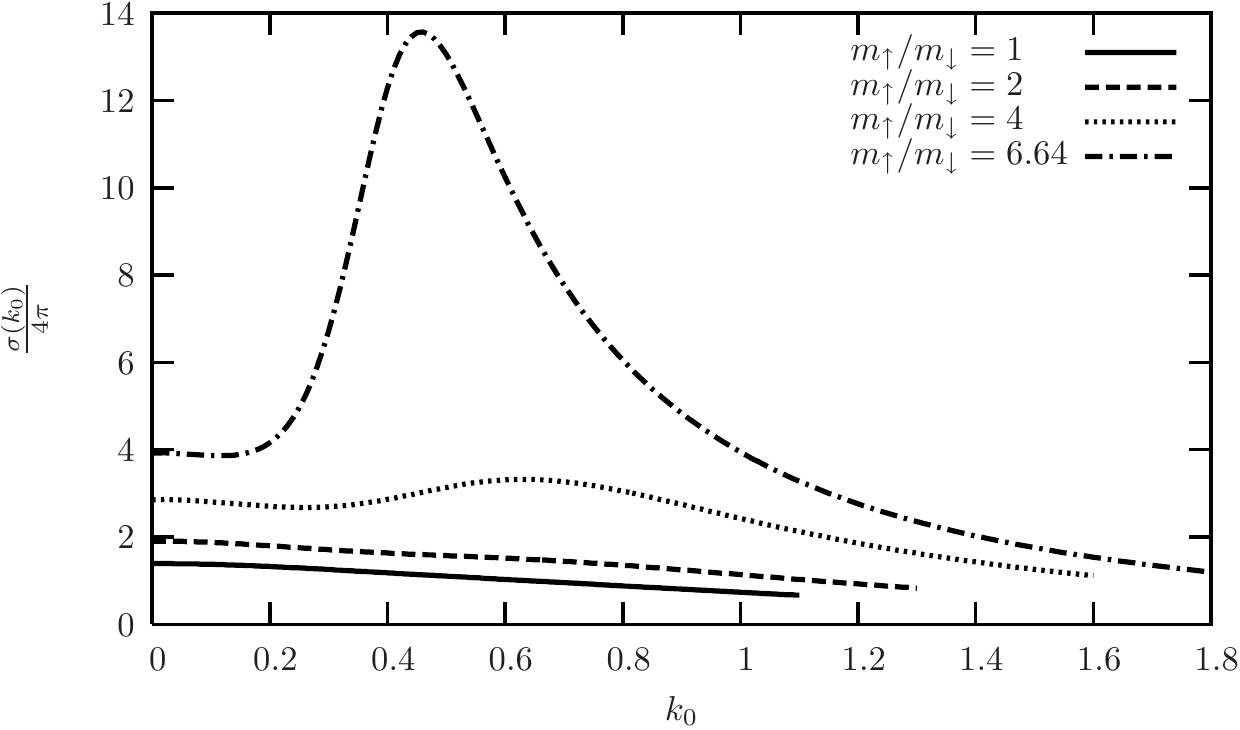}}
\caption{Total scattering cross section (divided by $4\pi $) Eq.(\ref{sigm}) as a function of $k_0$, for mass ratios
$m\us / m\ds=1, 2, 4$ and $6.64$.}
\label{fig5}
\end{figure}

\begin{figure}
\centering
{\includegraphics[width=\linewidth]{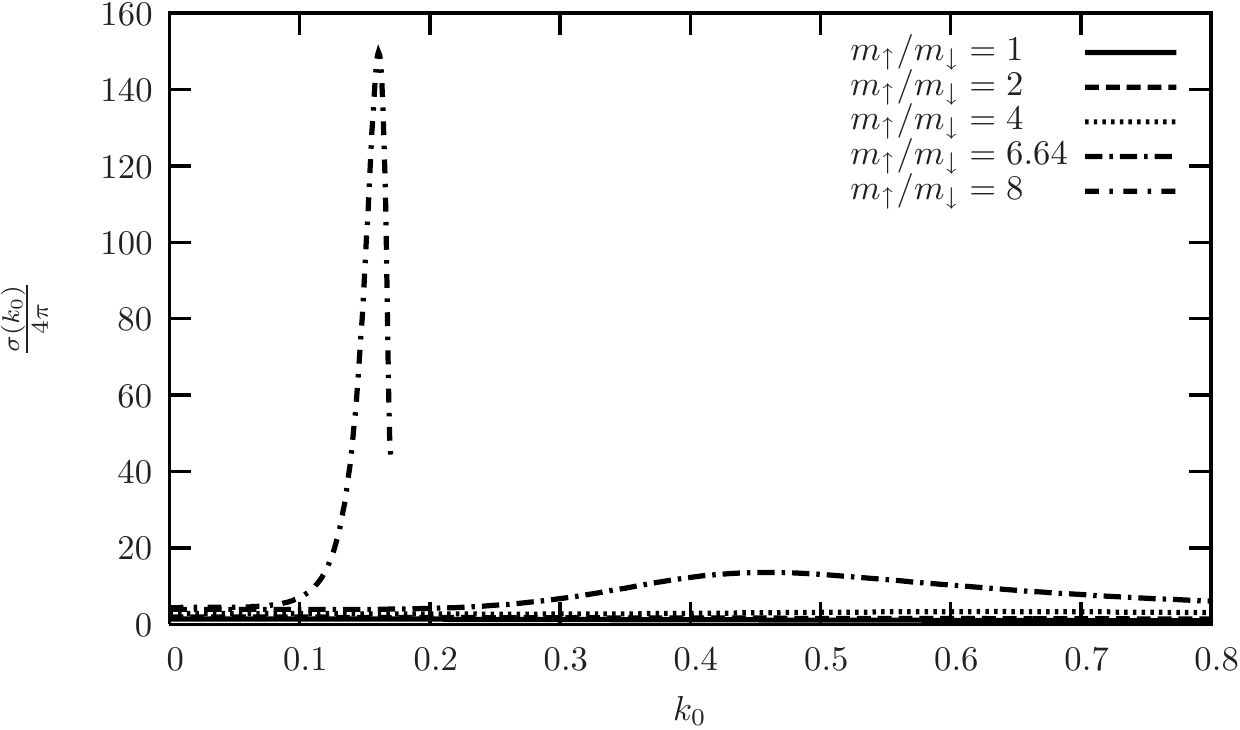}}
\caption{Total scattering cross section (divided by $4\pi $) Eq.(\ref{sigm}) as a function of $k_0$, for mass ratios
$m\us / m\ds=1, 2, 4$ and $6.64$ and $8$.}
\label{fig5bis}
\end{figure}

Let us now come to the total scattering cross section $\sigma(k_0)$, which is likely to be the easiest direct physical quantity to measure
experimentally. Since we neglect angular momenta higher than $\ell=1$, we have for this cross section:
\begin{eqnarray}\label{sigm}
\frac{\sigma(k_0)}{4\pi }=\!\sum_{\ell} (2 \ell +1)|f_{\ell}(k_0)|^2\!=\!|f_0(k_0)|^2+3|f_1(k_0)|^2
\end{eqnarray}
The result is plotted in Fig.\ref{fig5} for $m\us / m\ds=1, 2, 4$ and $6.64$. We see that, while up to $m\us / m\ds= 4$,
the cross section is almost featureless, it display a strong resonance for $m\us / m\ds=6.64$ which corresponds to the
experimental value for $^6$Li - $^{40}$K mixture. Naturally when the mass ratio is further increased, this resonance
becomes even stronger. This is shown in Fig.~\ref{fig5bis} where we display also the case of $m\us / m\ds=8$. The
resulting resonance dwarfs the preceding one.

This strong resonance in the atom-dimer scattering cross section for $^6$Li - $^{40}$K mixtures should imply a number 
of experimentally observable consequences. The more natural way
to evidence it would be to measure the collision properties between a $^{40}$K cloud and a $^{40}$K -$^6$Li dimer
cloud, in a way analogous to the recent experiments on $^6$Li for spin states \cite{zwierlein}. However this might not be so easy to perform.
Measuring the equation of state of a mixture is on the other hand a fundamental question which should be easier to answer.
The resonance should affect in an important way the low temperature equation of state of a mixture of $^{40}$K cloud and a $^{40}$K 
-$^6$Li dimer, arising from an unbalanced mixture of $^{40}$K and $^6$Li atoms on the BEC side of the Feshbach
resonance. This will not appear at the mean field level since the angular average will cancel the effect because this
is a p-wave resonance. But going to next order in $^{40}$K, analogously to second order perturbation theory, there will be no
cancellation. Hence one should see a strong dependence on the $^{40}$K density. One expects also a strong density
dependence to arise when the Fermi wavevector of the $^{40}$K Fermi sea reaches the wavevector corresponding
to the resonance, since for lower wavevectors the scattering is essentially negligible. One should see something
similar to a threshold effect in density. We note that the resonance should also have an important effect even on a
balanced mixture of $^{40}$K and $^6$Li, on the BEC side, when the temperature is raised. Indeed the dimers will
be partially broken by thermal excitation which provides a natural source of free $^{40}$K atoms, and effects analogous
to the ones arising in the unbalanced mixture. Similarly, going to high temperature, we expect this resonance to have
a marked effect on the virial coefficients since these are systematically related to the three-body problem \cite{xav}.
Naturally the resonance is also expected to affect strongly
the transport properties. Both the effects on the equation of state and on the transport properties should appear
in the frequency and the damping of the collective modes \cite{gps}.

Another way to understand the effect of this resonance on the $^6$Li - $^{40}$K mixtures is to remark that
the long lived virtual bound states responsible for it should behave in many respects in a way analogous to real bound states,
as it has already been noted by Levinsen \emph{et al} \cite{lp}.
Hence on the BEC side the physical description should involve not only free fermions and dimers, but
also the existence of trimers. Obviously all the physical properties should be affected by the presence
of this additional fermion species. A related point is that the existence of these trimers should affect the
dimer-dimer scattering properties. In other words we have only explored the simpler fermion-dimer
scattering properties, but we expect that the dimer-dimer scattering amplitude will display related
energy structure, and they should not be so complicated to explore with the methods of our preceding
work \cite{bkkcl}.

Finally it is worth stressing that the use of optical lattices should provide a very convenient and powerful way
to explore the resonance we have pointed out. Indeed the natural mass ratio of $6.64$ between $^{40}$K 
and $^6$Li atoms happens to be just at the border of the strongly resonating mass ratio domain. A rather weak
optical lattice could be used to slightly tune the effective mass of $^{40}$K or $^6$Li (by making an appropriate
choice of the light frequency, only one atomic species is affected), producing strong modification of the resonance 
and hence of the physical properties of the mixture. In this way one could go into the strongly resonating domain, 
or even reach the threshold for the appearance of real bound states. Or one could go in the other direction and basically
get rid of the resonance, which would allow to prove that it is responsible for specific physical properties of
the mixture. More specifically the 3D potential $V_{\rm opt}({\bf r})=s E_R \left[\sin^2(K_0x)+\sin^2(K_0y)+\sin^2(K_0z)\right]$
with $E_R=K_0^2/2m$ provides, within second order perturbation theory, an effective $m^*$ given by $m/m^*=1-s^2/32$.
Increasing the $^{40}$K effective mass to reach the bound state ratio $8,17$ requires $s=2.45$ which corresponds
to a fairly weak optical potential. This simple picture of effective mass modification works only if the size of the
involved objects (dimer, trimer) is large compared to the optical wavelength $\lambda$. Taking $\lambda \sim 500\, {\rm nm}$
this should be a valid approximation in the vicinity of the Feshbach resonance, 
where the scattering length becomes quite large. Even if corrections
to this simple picture are necessary, the qualitative physical trends should remain valid.

\section{Conclusion}

In this paper we have studied the s-wave and p-wave contributions to the fermion-dimer scattering amplitude as a function
of the mass ratio. When masses are equal the physical situation is quite simple. The p-wave contribution is completely
negligible, and the modulus of the s-wave contribution is essentially constant. Hence the energy dependence of the
scattering properties is inessential, and the fermion-dimer scattering length $a_3$ is enough to fully characterize these
properties.

When the mass ratio is increased the situation becomes much more complex, both for s-wave and for p-wave.
For the s-wave contribution we have obtained an analytical solution in the asymptotic limit of very large mass ratio.
This allows to have explicitly the behaviour of the s-wave scattering amplitude. We find that it displays a large number
of zeros. These are directly linked to the very rapid increase of the phase shift with energy, a zero appearing each time
the phase shift crosses a multiple of $\pi $. At low energy this rapid increase of the phase shift is directly linked to
the known large value of the fermion-dimer scattering length for large mass ratio, although we find that at higher energy
the phase shift saturates and goes back toward zero. In the range of mass ratio corresponding to possible experimental
situations, the numerical solutions display a behaviour in qualitative agreement with the asymptotic limit, although the
agreement is not quantitative which is not surprising. Interestingly we find that a zero is still present in the s-wave
scattering amplitude down to an experimentally reasonable mass ratio of $m\us /m\ds \simeq 15$.

For the p-wave contribution the behaviour gets rapidly quite complex as soon as the mass ratio has a sizeable value.
This is basically due to the appearance of a fermion-dimer bound state when the mass ratio reaches the value $m\us /m\ds = 8.17$,
and we have limited our study to mass ratios below this threshold. Even below this threshold the existence of the bound state
appears through the existence of virtual bound states at positive energy. These give rise to a resonance in the scattering amplitude.
This resonance gets stronger and goes to lower energy when the mass ratio is increased toward the limiting value of
$m\us /m\ds = 8.17$, at which the resonance corresponds to a divergence at zero energy. It is very interesting that the mass
ratio $m\us /m\ds = 6.64$, corresponding to the $^{40}$K - $^6$Li mixtures, much studied experimentally,
is at the border where the p-wave resonance becomes important. Roughly, below this mass ratio we find a fairly small bump
in the modulus of the p-wave scattering amplitude, which is fairly unimportant in the overall scattering properties and may be
omitted in a first approach. On the other hand above this mass ratio, the p-wave resonance grows very rapidly and becomes
the dominant feature of the scattering. We have stressed explicitly that the use of optical lattices could allow to vary
experimentally the mass ratio throughout this very interesting range of mass ratio. One could in this way disentangle the role
of this p-wave resonance in the physical properties of these $^{40}$K - $^6$Li mixtures, which are likely to be quite complex.
In the same way one could check if these virtual bound states play an important role in the instabilities of these mixtures.


\begin{references}
\bibitem{sademelo}This has been done also recently by Iskin and S\'a de Melo \cite{ism}
\end{references}
\end{document}